# SCMCI: Secured Click and Mortar Commercial Interaction

Mausumi Das Nath[1], Tapalina Bhattasali[2]

St. Xavier's College (Autonomous), Kolkata

`m.dasnath@sxccal.edu,tapalina@sxccal.edu`

**Abstract.** The wide spread of click-and-mortar model offers an opportunity to consider a universal commercial interaction method. However, issues like privacy protection resist the widespread acceptance. Traditional SET and SSL protocols are designed using Public Key Infrastructure (PKI) where extensive computations are carried out. Our aim is here to design a protocol to secure any type of commercial interaction for online platform, which also considers mobile platform. Therefore, our focus is on reducing heavy computations and making the overall procedure faster. A Secured Click and Mortar Commercial Interaction (SCMCI) protocol is proposed here to improve the performance of the commercial interaction procedure through replacing time consuming public key encryption and decryption algorithms by hybrid logic including the use of symmetric key. Comparative analysis has been done with traditional SET protocol using cryptool to prove the efficiency of the protocol.

## 1. Introduction

The participants in any e-commerce transaction are the consumers and the merchants. The benefits to the customers are round–the–clock operation, time savings and hassle-free transactions. Similarly, the merchants save on their finances as well as it reduces the time and effort to carry out any such commercial operations. This has been possible due to the Secure Socket Layer which is presently being used in electronic commerce and is proved to be steady access to the digital medium during the transmission. Yet, there are certain factors which cannot be met due to the underlying problems of information security. The security features of authentication, confidentiality, message integrity, and non-repudiation can be met to a certain extent by implementing the SET (Secure Electronic Transaction) protocol, particularly on electronic payment system. Recently, e-payment system brings the participants together with higher rate of efficiency, faster transactional speed and credibility, but the security problem is still an issue of concern to both the cardholders and the merchants.

As transactional security has drawn much attention among the researchers and practitioners, this article focuses on a proposed mechanism that could handle all the security parameters while carrying out any electronic transaction efficiently by the card users and the merchants without losing out any confidential data over the digital medium.

The organization of the paper is as follows: literature study is detailed in Section 2. Section 3 lays down the proposed work and comparative analysis and Section 4 finally gives the conclusion and future scope of the work.

## 2. Literature Review

Although the SET protocol is widely used in E-payment system, there are still certain security issues that has provoked the researchers and authors to propose some suitable measures to combat the disadvantages of the SET protocol.

A new protocol has been proposed [1] which offers a much more secure and an efficient E-payment protocol. It provides an extra layer of protection for the merchants as well as the consumers, i.e., the cardholders. Before the completion of the entire e-payment process, the cardholders were asked to enter their password, thereby authenticating themselves that they are the actual cardholder. The third party(Visa,masterCard) did not pay any role in the authentication process, but the same was carried out using the issuer security certificate. Another advanced protocol referred as Advanced Secure Electronic Payment Protocol have been proposed [2],where ECC(Elliptic Curve Cryptosystem with $F2^m$ not Fp),SHA (Secure Hash Algorithm) and 3BC (Block Byte Bit Cipher) instead of RSA and DES algorithm have been used. It has been noted that when ECC was combined with Block Byte Bit Cipher, it enhanced the security strength to a large extent. Moreover, the processing time was significantly reduced alongwith the hazard of stealing the private keys. A review[3] about transaction processing on ecommerce website was carried out using SET (Secure Electronic Transaction) protocol. It has been portrayed that SET being a very comprehensive security protocol, utilizes cryptography to provide confidentiality of information, ensured payment integrity, along with identity authentication. The parameter of message confidentiality and security were taken care by cryptography, digital certificate and authentication by SMS. First the report gave an introduction about ecommerce websites and then how to build it. Moreover, it later explained how SET worked and the components involved in it. Thereafter, the report gave out the required design and implementation of this protocol. It was seen that The SET protocol is popular worldwide but it still has some issues with its efficiency. Hence, the authors proposed a light weight version [4] of the said protocol (known as LITESET) which showed a fair amount of reduction in computational time and communication overhead. This has been performed by the usage of a new cryptographic primitive termed as signcryption. Again, it was observed that the security feature of non-repudiation was still not maintained efficiently using the SET protocol [5]. This was due to the fact of complexities involved and the increasing overheads. Thus the authors proposed an upgraded version of the SET protocol where a highly secure session key sharing technique has been used. The server will create a session key which will then be transported between the participating entities securely by encryption with public keys. After the session key was distributed, then the involved entities could perform secure online dialogue through the shared session key using AES symmetric key algorithm. This has helped to achieve the confidentiality, authenticity of the participants, integrity as well as non-repudiation. Moreover, it involved a low storage cost. On the other hand, TSET, a Token based Secure Electronic Transaction has been proposed by authors [6],which is an improvement over the basic SET protocol. Based on this, the end to end security has been taken care of by the concept of tokens. A copy of the token was retained which helped in detecting tampering of data. Customers were aware of the genuinity of the merchants before initiating a transaction. A grading mechanism was also being proposed along with this to improve the quality of transactional service. Thus, the TTP ensured trust , quaranteed if any discrepancy occured related to the product, the same would be either replaced or refunded within a specific time period. Moreover, TTP provided a transactional history which ultimately restricted the parties from denying any transactional event. Product related information was only known to the cardholder and the merchant. To secure strongly the confidential data, a multiple encryption technique have been proposed [7]. It was little different from the conventional SET algorithm. In the encryption phase, the process encrypted Message Digest several times with different encryption keys to make it more complicated and difficult to decode it back by the intruders. This newly generated Digital Signature was then transmitted to the receiver. The data security was so strong that no unauthorized user could even access any portion of the confidential information over the electronic medium. Still some researchers vouched that SET was safe [8], but claimed that the complexities must be removed and must be made simpler so that all the parties can adopt it to use it efficiently. But, the demerit was the cost involved in both hardware and software were quite high. Few authors developed a model checker to ascertain five essential accuracy properties of the Secure Electronic Transaction (SET) protocol [9]. This worked well even when the transacting participants, might have behaved fraudulently. An intermediary concept has been introduced[10] in a non-repudiation protocol. A complicated framework has been depicted using only one recipient followed by a situation where multiple recipients were present but mutual understanding between them was possible. The originator of the transaction was kept confidential to the receivers and vice-versa. The intermediate party might be distrusted but the security features were maintained in a non-reputable e-commerce event. An overview of Online Transactions and its Security were depicted along with the various security parameters and the techniques involved have been discussed [11].Conventional security techniques like SSL encryption, authentication does not always yield security. Thus, measures like, Anti key logging, biometrics, WebPin Technology and SET protocols have been adopted to give satisfactory security results. Few authors proposed [12] that popular identification methods like passwords have been replaced by biometric features of an individual. They have designed a new algorithm which had combined the biometric security, conventional technique, global positioning system and the mobile messaging. The reliability increased and the introduction of a three layer security model has enabled to boost up the security parameters of any electronic transaction. On the other hand, an approach was provided in [13] for securely authenticating the identity of a user who was a participating entity in an electronic transaction. Security for any device based transactions was also taken care of. Hence, researchers implemented a system [14], which enhanced security for device based transactions.

Again, authors proposed [15] that the use of encapsulated security token(referred as EST) can be strongly secured by applying functional data in one or more digital transactional event.EST was generated by encapsulating digital data including the operational data using at least two cryptographic systems of two users involved in the transaction. The encapsulation followed by de-encapsulation could use the cryptographic systems of the involved parties that involve a private key for signing and decryption, and a public key for encryption and signature verification .If the entire system could be formulated meticulously over a chain of painstaking events, the resulting EST can be practically difficult to forge .Moreover, a propagation of rights can be traced for auditing and rights can be easily concluded or modified.

However, in some scenarios it has been observed that computational cost was high, complexities have not been considered and also customer authentication has not been carried out.

## 3. Proposed Work

Here our aim is to design a secure and efficient protocol to protect any online shopping application from the fradulant activities. Our proposed protocol SCMCI (Secured Click and Mortar Commercial Interaction) focuses on major security dimensions like authentication, confidentiality, integrity, and non-repudiation for any click and mortar model. This model is designed based on the interactions among Customer(C), Merchant (M), Payment Gateway(PG) and Customer's Bank (CB), Merchant's Bank (MB) and Certificate Authority (CA). Initially, C, M, PG, CB and MB register themselves to CA to obtain Digital Certificates (DC).

### 3.1 SCMCI Procedure

*Step 1: Exchange $DC_n$ among set of participants (C, M, PG, CB, MB)*
*Step 2: Generate Symmetric Keys ($SK_n$) for each pair of participants to*
    *initiate commercial interaction.*
    *$SK_1(C,M)$: key between Customer and Merchant*
    *$SK_2(M,PG)$: key between Merchant and Payment Gateway*
    *$SK_3(CB,PG)$: key between Customer's Bank and Payment Gateway*
    *$SK_4 (MB, PG)$: key between Merchant's Bank and Payment Gateway*
    *$SK_5 (C, PG)$: key between Customer and Payment Gateway*
*Step 3: Generate $DE(SK_1(C,M), PUBK_C)$ // DE → Digital Envelope and*

$PUBK_C \rightarrow$ Public Key of C mentioned in $DC_C$

*Step 4:* Transmit $DE(SK_1(C,M), PUBK_C)$ from M to C.

*Step 5:* Generate $DE(SK_2(M,PG), PUBK_M)$ // $PUBK_M \rightarrow$ Public Key of M mentioned in $DC_M$

*Step 6:* Transmit $DE(SK_2(M,PG), PUBK_M)$ from PG to M

*Step 7:* Generate $DE(SK_5(C,PG), PUBK_C)$

*Step 8:* Transmit $DE(SK_5(C,PG), PUBK_C)$ from PG to C

*Step 9:* Generate $DE(SK_3(CB,PG), PUBK_{CB})$ // $PUBK_{CB} \rightarrow$ Public Key of CB

*Step 10:* Transmit $DE(SK_5(C,PG), PUBK_C)$ from PG to CB

*Step 11:* Generate $DE(SK_4(MB,PG), PUBK_{MB})$ // $PUBK_{MB} \rightarrow$ Public Key of MB

*Step 12:* Transmit $DE(SK_4(MB,PG), PUBK_{MB})$ from PG to MB

*Step 13:* Generate lookup table for set of participants (C, M, PG, CB, MB)

*Step 14:* Generates H(OS) and H(PD) for C // H(OS)$\rightarrow$hash of order summary and H(PD)$\rightarrow$hash of purchase details

*Step15:* Calculate $Cipher_C \leftarrow enc(SK_1(C,M)(H(OS)))//enc(SK_5(C,PG)H(PD))$

*Step 16:* Transmit $Cipher_C$ from C to M

*Step 17:* Compare $Cipher_C$ with $Cipher_{C'}$ ( calculated at Merchant site)
//ensure message integrity, privacy, authenticity, and non-repudiation, since $SK_1(c,m)$ only known to C and M
//at the time, when merchant receives the message, the following steps are performed as follows:

*Step 18:* Calculate $dec(SK_1(C,M)(enc(SK_1(C,M)(OS)))$ to get OS

*Step 19:* Calculate $dec(SK_1(C,M)(enc(SK_1(C,M)(H(H(OS) || H(PD))))$ to get $H(H(OS)//H(PD))$
//after merchant receives the OS message, merchant computes the following:

*Step 20:* Compute H(OS) and $enc(SK_2(M,PG)H(OD))$

*Step 21:* Transmit $enc(SK_2(M,PG)H(OD))$ and $enc(SK_5(C,PG)H(OD))$
// PG verifies OD message is not altered during processing
//PG verifies PD message as follows:

*Step 22:* Transmit PD from PG to CB // through a secured and private extranet

*Step 23:* CB verifies PD // to verify ability of C for this interaction

*Step 24:* Transmit authorized information(AI) from CB to PG // through secured financial extranet.

*Step 25:* Sign authorization response(AR) and transmit from PG to M

*Step 26:* Debit amount at CB for C and transmit payment response

    *(PR) to PG*
*Step 27: Redirect PR from PG to MB*
*Step 28: Credit amount at MB for M //ensure consistency among the*
     *various İnteractions*
*Step 29: Transmit PR from PG to M for verification*
*Step 30: Transmits goods or services from M to C*

## 3.2 Analysis

We have analyzed the performance of proposed protocol quantitatively using Cryptool. It reduces the computational overhead compared to the traditional SET protocol. We have also analyzed the protocol qualitatively to prove the improvement of the proposed protocol from the existing protocols.

The inputted content is being encrypted with our proposed algorithm.First,symmetric key encryption has been used, then the hash value (using MD5) has been used to generate the encrypted message. Later,this message is again encrypted.Finally, the entropy value has been calculated.The same process has been performed by using asymmetric encryption technique.It is observed that the entropy value is less in our proposed method than the asymmetric method used.Hence, it is justified that our proposed technique is much faster, complextity is lower and the overhead is also reduced.

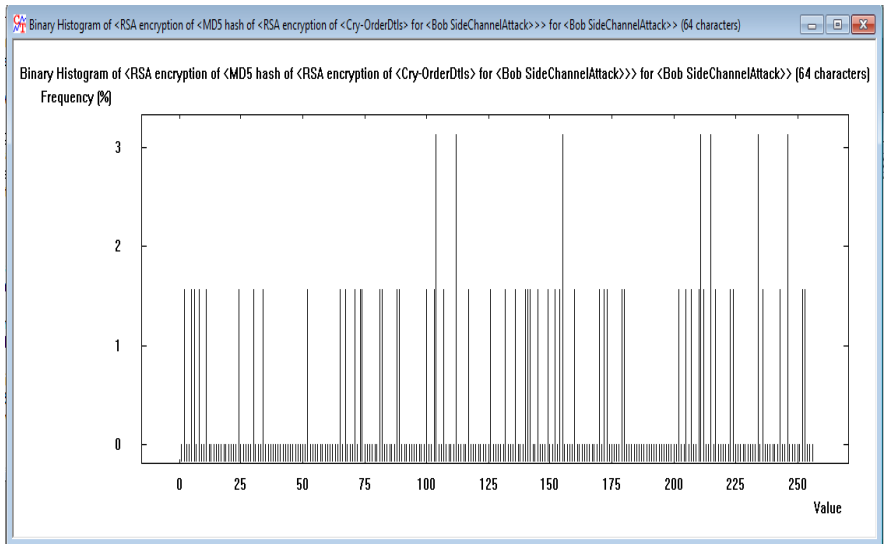

**Fig1:** Histogram of Order Information using Traditional SET Protocol
(Entropy-5.78)

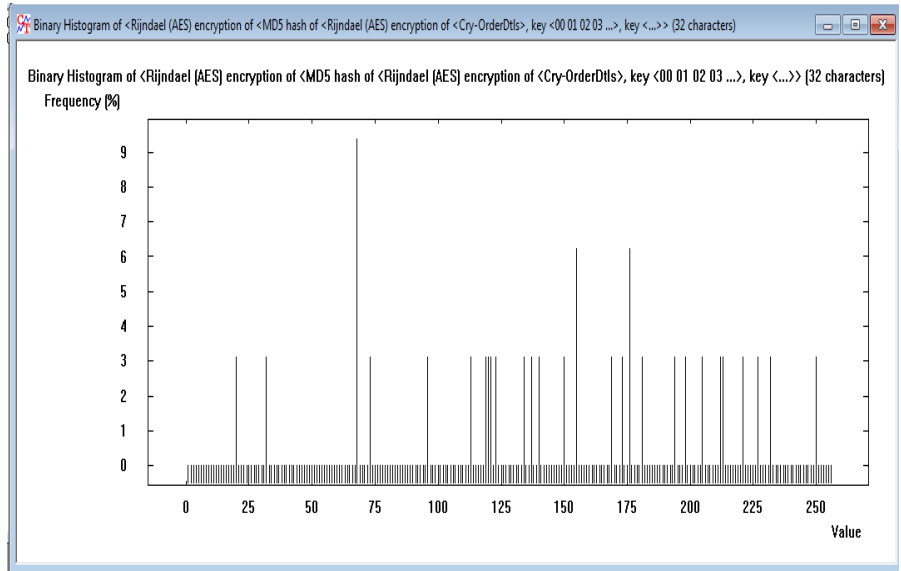

**Fig2:** Histogram of Order Information using Proposed SCMCI Protocol
(Entropy-4.72)

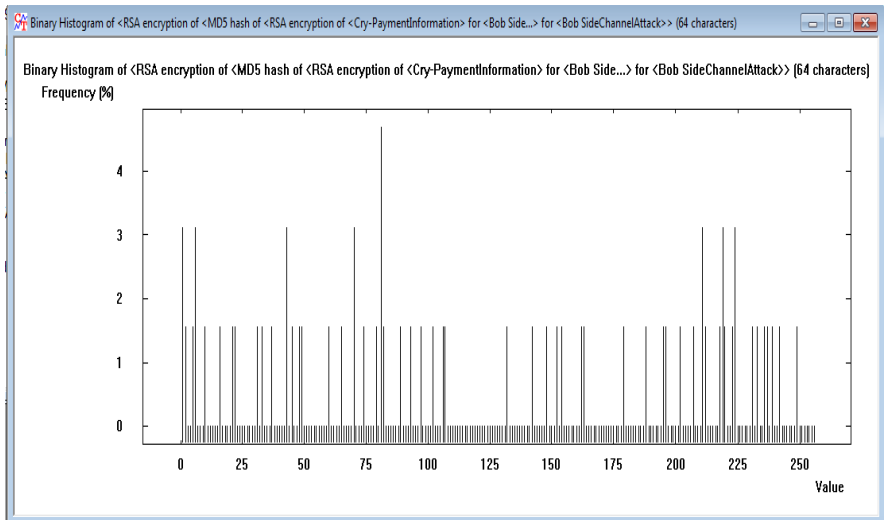

**Fig3:** Histogram for Payment Information using Traditional SET Protocol
(Entropy-5.70)

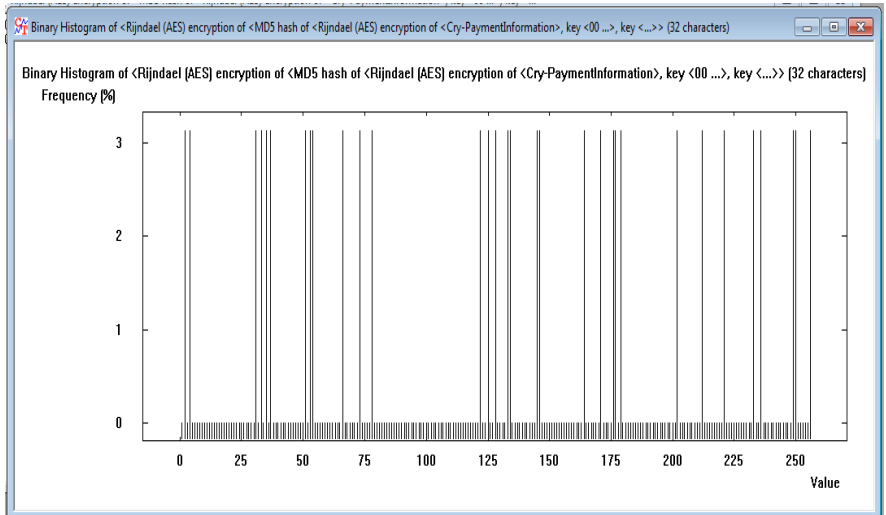

**Fig4: Histogram for Payment Information using Proposed SCMCI Protocol (Entropy-5.0)**

**Analysis of Side Channel Attack for traditional SET protocol**

Customer composes a message M, addressed to Merchant. Customer chooses a random session key S:
    FD68C116F776643E447A6EEDC77ECDFD

### MESSAGE ENCRYPTION

Customer symmetrically encrypts the message M with the session key S. Customer chooses Merchant's public key e: **010001** Customer asymmetrically encrypts the session key S with
Merchant's public RSA key e:
    09CAAED47E518ED0E93CB4D2B5D259C4B48578E054F051F7
    E1C025238AC9A3E5790F2AA87EDA93350644D7B4B4C2FB70
    60C5782AE05A0CE19A3F7A0406F32AF0

### MESSAGE TRANSMISSION

Customer sends the hybrid encrypted file to Merchant over an insecure channel.

### MESSAGE INTERCEPTION

Trudy intercepts the hybrid encrypted file and isolates the encrypted session key S:
    09CAAED47E518ED0E93CB4D2B5D259C4B48578E054F051F7
    E1C025238AC9A3E5790F2AA87EDA93350644D7B4B4C2FB70

60C5782AE05A0CE19A3F7A0406F32AF0

**BEGINNING OF THE ATTACK CYCLE**

Customer sends an exact copy of the original, encrypted message to Merchant and extends it with the session key S' (encrypted with Merchant's public key). Compared to the message sent by Customer, Trudy (an intruder) simply replaces the encrypted session key [ENC(S, PubKeyMerchant) is replaced by ENC(S', PubKeyMerchant)]. Trudy repeats this step 130 times, whereas the step count depends on the bit length of the used session key (step count = bit length + 2).

Repeating the above mentioned steps for our proposed protocol, it has been seen that our protocol is resistant to side-channel attack. Message cannot be intercepted by third party, which can enhance the trust of customers for click and mortar model.

## 4 Conclusions

Our proposed protocol SCMCI (Secured Click and Mortar Commercial Interaction) will play an effective role in shopping application by providing essential security
features such as confidentiality, integrity, availability, authentication and non-repudiation for any click and mortar model.It earns the trust of both the customers and merchants.The customer feels safe as their data is not intruded upon and the quality of service is also met.The intruder even after repeated attempts cannot succeed in intercepting the confidential data.Thus, the customer's data is safe and protected.Merchant can also easily verify customer's identity.The computational overhead is reduced,enables in faster transaction and is also trustworthy. In future, proposed protocol can be analyzed for other threats to prove the efficiency of the protocol. SCMCI can also be extended to implement it for any click and mortar interactions.